\documentclass[final,1p,times,twocolumn]{elsarticle}

\usepackage{amssymb}
\usepackage{amsthm}

\usepackage{lineno}

\journal{New Astronomy}

\begin{document}

\begin{frontmatter}



\title{The First Gravitational-Wave Burst GW150914, as Predicted by the Scenario Machine}


\author{V. M. Lipunov$^{1,2}$, V.Kornilov$^{1,2}$, E.Gorbovskoy$^{2}$, N.Tiurina$^{2}$, P.Balanutsa$^{2}$, A.Kuznetsov$^{2}$}
\address{$^{1}$M.V.Lomonosov Moscow State University, Physics Department, Leninskie gory, GSP-1, Moscow, 119991, Russia \\
$^{2}$M.V.Lomonosov Moscow State University, Sternberg Astronomical Institute, Universitetsky pr., 13, Moscow, 119234, Russia\\}
\ead{lipunov2007@gmail.com}
\begin{abstract}
The Advanced LIGO observatory recently reported ~\citep{Abbott2016a} the first direct detection of gravitational waves predicted by Einstein (1916) ~\citep{Einstein}. The detection of this event was predicted in 1997 on the  basis of the Scenario Machine  population synthesis calculations ~\citep{Lipunov1997b}
Now we discuss the parameters of binary black holes and event rates predicted by different scenarios of binary evolution. We give a simple explanation of the big difference between detected black hole masses and the mean black hole masses observed in of X-ray Nova systems. The proximity of the masses of the components of GW150914  is in good agreement with the observed initial mass ratio distribution in  massive binary systems, as is used in Scenario Machine calculations for massive binaries.
\end{abstract}
\end{frontmatter}

\section{Introduction}

The Advanced LIGO observatory recently reported the first direct detection of gravitational waves  by the merging of two black holes with  masses of $36^{+5}_{-4} M_{\bigodot}$ and $29^{+4}_{-4} M_{\bigodot}$ \citep{Abbott2016a,Abbott2016b,Abbott2016c}. In this article we  concentrate on the theoretical interpretation of GW150914 in the framework of binary population synthesis.
The fact, that some double stars have relativistic components, has never been doubted since the publication of the first evolutionary scenario of massive binaries ~\citep{Clark,Paczynski,Tutukov1973,Heuvel} which before producing two relativistic stars are expected to spiral in to a very narrow binary system ~\citep{Heuvel73}. This was confirmed by the direct discoveries of such systems in our Galaxy \citep{Hulse}. The study of such systems excellently demonstrated the correctness of general relativity including the predicted loss of angular momentum of a double radio pulsar exactly in accordance with Einstein’s formula \citep{Einstein}. Observations of the binary pulsar PSR 1913+16 showed that the merging time of such systems is shorter than the Hubble time and this fact formed the basis for the first experimental estimates of the double neutron star merging rate in the Universe and the probability of finding this process ~\citep{Clark,Phinney}.

On the other hand, only in the early 1980s years the first method of population synthesis calculations of late stages of binary stellar evolution study  was developed. The developed codes took into account  the formation and evolution of relativistic stars including neutron stars and black holes ~\citep{Kornilov1983a,Kornilov1983b}.
 This method made it possible to determine for the first time the expected double neutron star merging rate normalized to the constant star-formation rate characteristic of our Galaxy with the mass of $10^{11} M_{\bigodot }$ \citep{Lipunov1987} and compute the amplitude and continuum spectrum of the gravitational-wave background produced by \textbf{double} black holes \citep{Tutukov1993,Lipunov1995}.
  However, only in 1997 it could be shown that for  almost all binary scenarios input parameters the first events to be recorded by LIGO type detectors should be the merging of relativistic systems with black-hole components: BH+BH or BH+NS  \citep{Lipunov1997a,Lipunov1997b,Lipunov1997c}.

In this paper we describe the theoretical part of the MASTER project, which concerns the calculation of the predicted GW event rate to be detected by LIGO/VIRGO.
The MASTER optical follow-up observations of GW150914 as a part of its electromagnetic investigations will be  discussed in reference \citep{MASTER_LIGO_Observ}.

\label{}

\section{Scenario Machine prediction and explanation of the GW150914 event}

\begin{figure*}
 \centering
 \center{\includegraphics[width=0.7\linewidth]{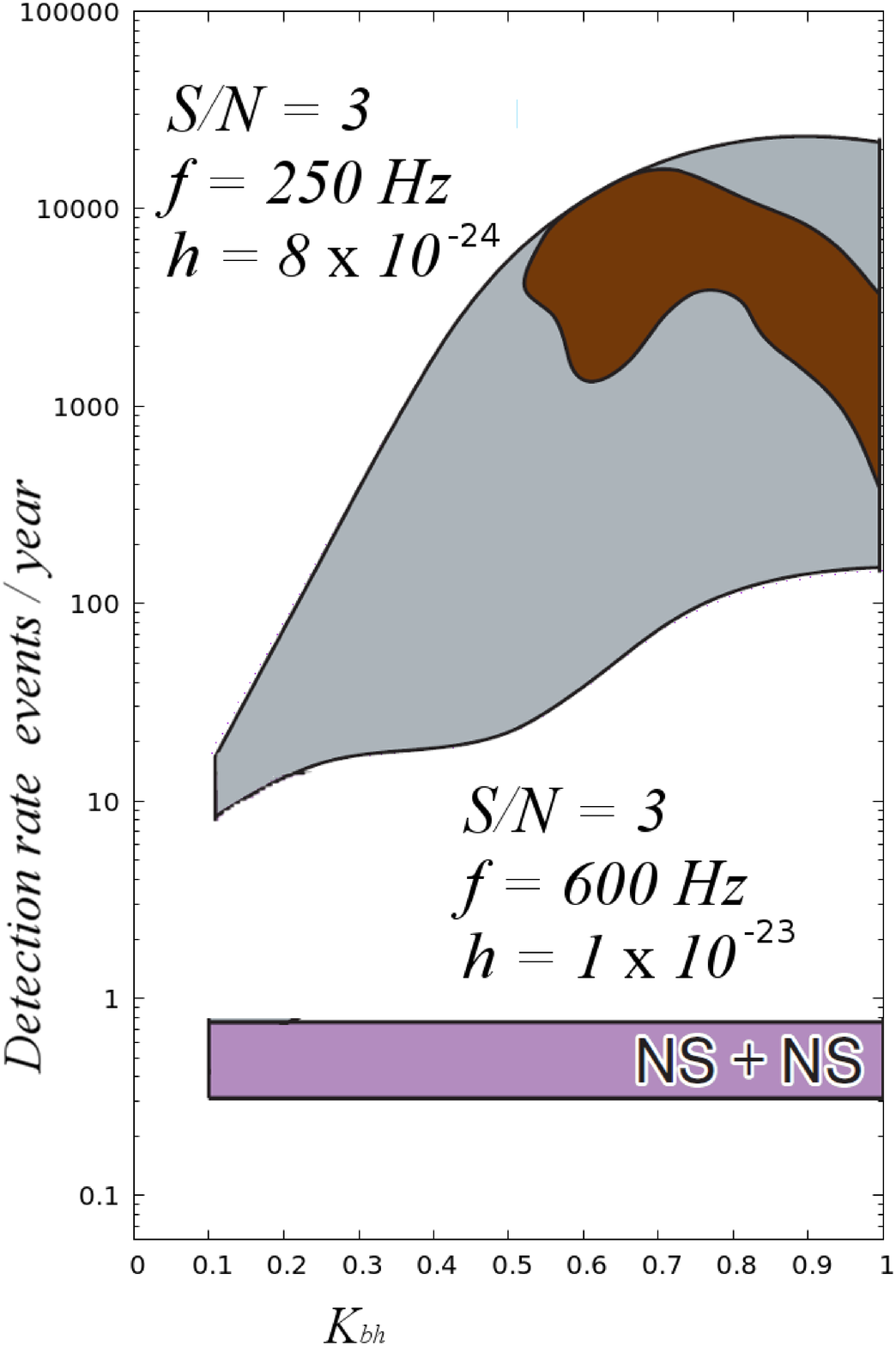}}
 \caption{
 Updated version of  Figure 3 of paper ~\citep{Lipunov1997a} for modern LIGO parameters. We recalculated the detection rates for sensivity
 $ h_{\nu} = 8 \cdot 10^{-24}$ at  $ 250 Hz$ for BH + BH and BH + NS collisions and
 $ h_{\nu} = 1 \cdot 10^{-23}$ at  $ 600 Hz$ for NS + NS.
 The expected detection rate of gravitational-wave bursts from neutron star and black hole mergers as a function of the unknown parameter $k_{BH}$, the fraction of the star’s mass carried into the black hole at the time of its formation (for $NS+NS$ systems $k_{BH}$ has no relevance; here just the predicted rate is plotted). The dark area
 shows the likely detection rate domain computed based on modern models of the evolution of binary stars. The size of this domain is very large because many parameters are unknown, however, it is everywhere above the detection rate for signals from neutron-star mergers (the NS+NS strip). The diagram shows that the first  gravitational-wave sources to be discovered will be merging binary black holes Lipunov et al. \citep{Lipunov1997a}
 }
 \label{FigureGolova}
 \end{figure*}

Lipunov et al. in ~\citep{Lipunov1997a} showed that irrespective of the particular evolutionary scenario and the initial binary parameters  the first events, that will be detected with LIGO types of interferometers, should be   black holes mergers. This result was most clearly  demonstrated in Figure 3 of paper ~\citep{Lipunov1997a},  which we recalculated here using the real sensitivity of the LIGO detectors on 14 september 2015 (Fig. \ref{FigureGolova}).

The first updates are connected with the Signal to Noice ratio. In reference ~\citep{Lipunov1997a} this figure was calculated for  Signal to Noice ratio $S/N_{old} = 1$. Here we recalculate if for  $S/N_{new}  = 3$, which  is closer to the threshold of detection used in the present LIGO. The second change is associated with  the recalculation for the real sensitivity of the LIGO detectors on 14 September 2015. We use  the strain noise of the LIGO detector (Figure 3b in ~\citep{Abbott2016c}) as sensitivity $h_{\nu}$. In our calculation we use  the dimentionless sensitivity $h_{rms} = h_{\nu} * \sqrt{\nu}$. The GW150914 event was detected at frequencies from 35 to 250 Hz. Now we  use  $ h_{\nu}^{(bh)} = 8 \cdot 10^{-24}$ at  frequency  $ 250 Hz$  (at which GW150914 was observed) for BH + BH and BH + NS collisions. Consequently today the dimentionless sensitivity  $ h^{(bh)}_{rms} = 1.26 \cdot 10^{-22} $ should be used  instead of  $ h_{rms}=1 \cdot 10^{-21} $  , that was used in our 1997 papers \citep{Lipunov1997a,Lipunov1997b,Lipunov1997c}. As a result the rate of detection should increase at $ (\frac{h_{old}}{h_{new}})^3  \times  (\frac{S/N_{old}}{S/N_{new}})^3 \sim 18.9  $. The same calcultion for   (NS + NS) collision with $ h_{\nu}^{(ns)} = 1 \cdot 10^{-23}$ at  frequency  $600 Hz$  leads to  $ h^{(ns)}_{rms} = 2.45 \cdot 10^{-22} $ . This increases the rate by $ \sim 2 $ times for NS + NS coalescences.



The point we made earlier is that the most likely objects to be detected by gravitational-wave detectors are  massive binaries for which the initial masses of both components are substantially greater than 10 solar masses. These are the systems, which are capable to produce the binary black holes or the binaries consisting of a black hole and neutron star at the end of their evolution.

Let us discuss this type of systems, believing the first event is a merging of binary black holes.


The occurrence rate can be estimated analytically only with an uncertainty of several orders of magnitude. That is why a special method of population synthesis based on Monte-Carlo technique – i.e. the Scenario Machine ~\citep{Kornilov1983a,Kornilov1983b} - was proposed for analyzing various scenarios of the evolution of binary stars.

 The very first Scenario Machine computations made it possible to determine the statistical properties of different types of massive binaries including the final stages of the stellar evolution involving the formation of relativistic binaries, which are the potential sources of gravitational-wave pulses at the time of their merging. The computations performed with the upgraded Scenario Machine allowed the occurrence rate of double neutron-star merger in our Galaxy for the given star-formation rate and Salpeter initial mass function to be determined as early as 1987 ~\citep{Lipunov1987}.

In 1993 Tutukov and Yungelson ~\citep{Tutukov1993} performed the first computations of rate of black hole mergers, and showed that the merging rate of such systems can be comparable to that of neutron stars. However, the very large number of poorly determined parameters of binary star evolution prevented the understanding of the actual occurrence rate of events recordable by LIGO/VIRGO type gravitational-wave detectors.

The most successful attempt was performed by ~\citep{Lipunov1997a}, who used the Scenario Machine to this end. Note that unlike other types of population synthesis the Scenario Machine is designed to compare the results of numerical simulations with the entire available set of observational data about relativistic stages of binary star evolution: radio pulsars in binaries with components of different types, x-ray pulsars, black-hole candidates, and millisecond radio pulsars.

As a result, the Scenario Machine made it possible to adjust the parameters of stellar evolution so that the predicted distribution of neutron stars and black holes in binaries would be consistent with the observations. We even allow the possibility that this approach, which involves so many parameters, can be used to obtain reasonably credible predictions about hitherto unobserved processes.

We list here the main parameters ~\citep{Lipunov1996b}: the distribution function of component mass ratios $\varphi(q) =M_2/M_1 <1)$; the efficiency of the common envelope evolution, the kick velocity distribution during relativistic star formation, stellar wind mass loss (high mass loss and low mass loss scenario). And  other, better  determined parameters, such as the Salpeter initial mass function and the distribution of the semi-axes of binary stars.

To obtain the most reliable prediction concerning the first events to be detected with gravitational-wave interferometers, Lipunov et al. ~\citep{Lipunov1997a} computed a scenario with weak stellar wind varying all the poorly known parameters mentioned above. The only concern was to ensure that there would be at least one Cyg X-1 type system (Black hole with a blue supergiant) and no pulsars with a black hole companion per 1000 single radio pulsars ~\citep{Lipunov1994,Lipunov2005}. Notice that so far no such systems have been discovered although about 2000 single radio pulsars have been found.

The first condition evidently imposes a lower limit on the double black hole merging rate, whereas the second condition provides an upper limit for this parameter. The large grey domain in Figure \ref{FigureGolova} is the result of our computations for all possible values of the parameters mentioned above. It is evident that the scenario with a weak stellar wind leads to a clear result: the first events to be detected on LIGO/VIRGO interferometers undoubtedly should involve black holes! We also came to the same conclusion in our analysis performed in the framework of a scenario with strong mass loss in the form of stellar wind ~\citep{Lipunov1997b,Lipunov1997c}.

What does the large mass of the First Gravitational Wave Burst  GW150914 imply? The first evident conclusion is that this event was a result of the evolution of a very massive binary having weak stellar wind, i.e., the scenario that we used in ~\citep{Lipunov1997a}.

\begin{figure*}
\centering
\center{\includegraphics[width=0.6\linewidth]{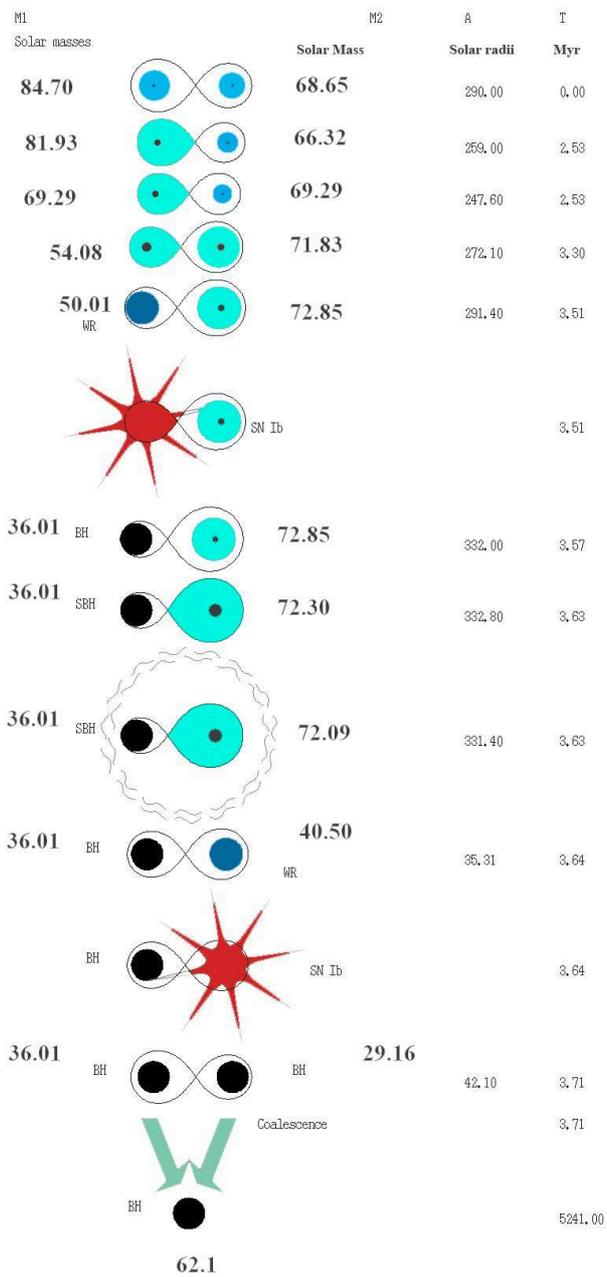}}
\caption{Binary evolution track for GW150914 generated by the online Scenario Machine code (Nazin et al., 1998 http://xray.sai.msu.ru/sciwork/scenario.html ).}
\label{FigureScenario}
\end{figure*}

In Figure \ref{FigureScenario} we depict a possible evolutionary track resulting in the merging of two black holes with the masses of 29 and 36 $ M_{\bigodot} $ (the  initial system parameters and evolutionary assumptions of this scenario are listed in Table \ref{table_param}). This evolutionary sequence was generated with online Scenario Machine \citep{Nazin} \footnote{http://xray.sai.msu.ru/sciwork/scenario.html} ; one sees that the system undergoes two supernova explosions and a common-envelope phase; we assumed a weak stellar wind. Already in 1997 we generated several such evolutionary sequences, and the computations by Belczynski et al ~\citep{Bel16} produced very similar evolutionary sequences for explaining GW150914, confirming the earlier results of Lipunov et al. \citep{Lipunov1997a,Lipunov1997b,Lipunov1997c}.
We started calculations for low metallicity
stars in these papers, but the effect of low metalicity does not dramatically change the results
 in the Scenario Machine the assumed logarithmically flat initial separations distribution of binaries appears to largely compensate for the effects of low mass loss rates on the evolutionary tracks.

 We published 3 papers: for low mass loss (low metallicity ~\citep{Lipunov1997a}), high mass loss and different kick velocity \citep{Lipunov1997c} and in \citep{Lipunov1997a}).
 With the Scenario Machine we simulated all binary stages with a relativistic companion , i.e. binary radio pulsar, Cyg X-1 type, X-ray pulsars (accreting neutron stars), accreting white dwarfs (Cataclysmic binaries), binary radio pulsars with massive OB star companion,  etc., and we adapted the key input parameters of the code such that they produced best fit of the output with the observations of these different types of objects. Other of population synthesis codes did not follow such a procedure of check of output with all observed types of relativistic binaries anywhere up to now, because nobody included the calculation of the spin evolution relativistic stars.

Although the computations reported by Lipunov et al.~\citep{Lipunov1997a} did not imply a scenario of the evolution of Population-III binaries
~\citep{Kinugawa},
the results obtained are applicable to Population III because in our computations stellar wind does not carry away any significant fraction of the progenitor mass, as is typical of stars with low content of heavy elements. On the other hand, \citet{Lipunov1995} did perform  computations of the evolution of the merging rate of relativistic stars that took first-generation stars into account. We showed that the rate of neutron stars merging with black holes (Figure \ref{FigureGolova}, \citet{Lipunov1995}) decreases by 3.5 orders of magnitude after the second billion years, and reaches a constant level of

$$
R_{M}  \sim  \frac{ 10^{-7} yr^{-1} }{ 10^{11} M_{\bigodot} } \eqno(1).
$$

To convert this quantity into the merging rate per unit volume in the comoving frame, we have to use the data about baryon density involved in star formation (see formula 5 in \citep{Lipunov1997b}):

$$
R_{V} =  R_{M} \; (\varepsilon/0.5) \; (\Omega_{b}/0.0046) \; H_{75} \;  M pc^{-3} \eqno(2)
$$

Where $\Omega_{b}$  - is the baryon density of the luminous matter in the Universe (in units of the critical density),  $\varepsilon$ - fraction of the baryons in  binary stars (typically adopted as $0.25 < \varepsilon < 0.75$) and $H_{75} = H \; / \; 75 \; \frac {km /s}{Mpc^3} $  .  Substituting the average value $\epsilon=0.5$ we obtain a $100 \; Gpc^{-3} \;  yr^{-1}$  estimate for the current rate of merging of double black holes produced by first-generation stars. It therefore cannot due to ruled out that the progenitor of the observed event could be a first-generation stars.

\begin{table*}
 \centering
  \caption{ Track initial parameters \label{table_param}}
  \begin{tabular}{@{}lc@{}}
  \hline
Parameter with Description  & Value            \\
 \hline
A - semimajor axis $R_{\bigodot}$ &290 \\
M1 - primary mass, $M_{\bigodot}$ & 84.70\\
 M2 - secondary mass, $M_{\bigodot}$ & 68.65\\
E - orbital eccentricity   & 0.5 \\
 Normal star mass loss rate:   & Low\\
 Maximal accretion rate into CE:& Eddington\\
  Matter acception by normal star during accretion:    & Partially non conservative\\
    common envelope efficiency    & 1\\
      Minimal pre-SN mass for Black Hole  formation $M_{\bigodot}$ & 25\\
       collapse mass fraction (kBH)  & 0.8\\
       Oppenheimer-Volkoff limit $M_{\bigodot}$ & 2.5 \\
       initial spin period NS:   & $10^{-3}$\\
  \hline
\end{tabular}
\end{table*}

First, the event rate predicted by \citep{Lipunov1997a,Lipunov1997b,Lipunov1997c} is highly uncertain because of the lack of precise knowledge of the parameters of the evolution of binary progenitors of black holes (the uncertainty range amounts to almost three orders of magnitude). Therefore a comparison with the absolute merging rate in space and detector data requires a correct account of the sensitivity curve and actual noise of the detector. The signal amplitude is  about $1 \cdot 10^{-21}$, whereas the noise level is much lower, something about $0.2\cdot10^{-21}$ according to estimates based on the mean square deviation.

Furthermore, there is also a pure selection effect, which is responsible for the large total mass of black holes. The density of events with amplitude \textit{h} recorded by the detector can be estimated by considering a spherical shell around the observer of radius \textit{r} and thickness \textit{dr}. It is evident that

$$
dN (r/h_{0}) = 4\pi r^{2} \; dN(h_{0}) \; dr   \eqno(3)
$$

where $dN(h_{0})$ is the number of merging systems with gravitational-wave amplitude  $h_{0}$ per unit volume at unit distance. In terms of the observed amplitude  $h=\frac{h_{0}}{r}$ we obtain

 $$
 dN(h|h_{0}) = 4 \pi dN(h_{0}) \; \frac{h_{0}^{3}}{h^{4}} \; dh     \eqno(4)
 $$

 The final  distribution of gravitational-wave amplitudes recorded by the detector can be determined by integrating over all  $h_{0} = \Gamma M^{5/6}$  or over all chirp masses M

 $$
 dN(h) = \frac{4 \pi}{h^{4}} \int dN (h_{0}) h_{0}^{3} \; dh_{0}  = \frac{4 \pi}{h^{4}} \frac{5}{3}\Gamma ^{4} \int dN (M) M^{7/3} dM,     \eqno(5)
 $$
 where $\Gamma \sim 0.3 $ is dimensionless coefficient, which connects the beginning amplitude of gravitational wave  with chirp mass.

The  number  $N(h>\Pi) $ of events on the detector having an amplitude above certain threshold $\Pi$  is

$$
N(h>\Pi) = \frac{20 \pi}{\Pi^3} \Gamma ^{4} \int dN (M) M^{7/3} dM    \eqno(6)
$$

We illustrate these considerations by numerical computations made with the Scenario Machine. Figure \ref{figureDistrib} shows the distribution of the total mass of merging black holes in the present Universe and the corresponding distribution for the events to be recorded by the detector. As is evident from the figure, the median of the distribution shifts almost by a factor of two toward larger masses, and merging of double black holes with a total mass of $\sim 60$  appear to be nothing unusual.

Thus the discovery of a merging double black hole  by LIGO confirmed our view of the evolution of the most massive binaries.



\section{Discussion}

Why did the black holes turn out to be much more massive than expected?

The anomalously high masses of the merging black holes (in the opinion of many researchers) have since been discussed debated  during many discussions that had taken place after the discovery of  black-holes merging.
Indeed, according to the statistics of the  black hole candidates discovered in X-ray Novae binary systems, the average mass of the black hole is 
about 6-7 solar masses  \citep{Xbook,Cherep}.
 In most of these systems the optical component is a dwarf star of about solar or subsolar mass.
And such systems do not produce binary black holes that can be direct progenitors of LIGO events. As  pointed out above, double black holes are the result of the evolution of  systems where both components are  massive stars capable to produce black holes . In view of this fact we have to explain why the black holes in x-ray novae systems have low masses rather than why the black holes in GW 150914 are so massive.

Let us now return to the question why are the black hole masses in low-mass binaries
 so small. The point of view is that in the systems with the initial mass ratios
$ q_0 = M_2 / M_1 \lesssim 100 $
 the dwarf star simply has no time to reach the main
sequence and is evaporated  by its millions of times more luminous blue companion.

 One can easily show that the black holes in X-ray binaries with an about $1 M_{\odot} $ donor star cannot have been produced by companion stars that started out much more massive than about $25 M_{\odot} $; such stars do not produce black holes more massive than about $10 M_{\odot} $.

 At the same time, the relatively large mass of the GW150914 events is quite consistent with the parameters of the evolutionary scenario of massive stars with weak nuclear stellar wind computed with the allowance for selection effects due to the increase of merging detectability horizon with the total mass of the binary system.

There is yet another important circumstance suggested by the parameters of the GW150914 event. It is the fact that the merging black holes have rather similar masses.
We calculated the final mass ratio distribution for two cases:$\varphi(q) \sim q^{2}$ as proposed in  ~\citep{Tutukov1985}, and  $\varphi(q) = const$ as indicated by ~\citep{Sana}. The result is not dramatically changed.


 With the assumption \textbf{$\phi(q) ~ \sim q^2$} , in 90\% of merging  double black holes are  black holes with the similar masses.

This paper is the theoretical paper concerning the MASTER project for the theoretical and practical study of the gravitational-wave event GW150914.
The MASTER optical follow-up observations of GW150914 as a part of its electromagnetic investigations will be  discussed in  reference \citep{MASTER_LIGO_Observ}.

\section{Acknowledgement}
One of us (VL) would like to thank Kip Thorn and remember  Leonid Grishchuk, the person who predicted the cosmological gravitational wave background ~\citep{Grishchuk1977}. It is the suggestions of these people that prompted us to perform the corresponding computations.

The MASTER project is supported in part by the Development Programm of Lomonosov Moscow State University, Moscow Union OPTICA,  Russian Science Foundation 16-12-00085, Russian Foundation of Fundamental Research 15-02-07875

We are especially  grateful to S.M.Bodrov for his long years MASTER's support.

The author are gratefull to Prof. Edward Van den Heuvel and to referee for the very fruitfull discussion and remarks.


\begin{figure}
\begin{minipage}[h]{1\linewidth}
\includegraphics[width=0.7\linewidth]{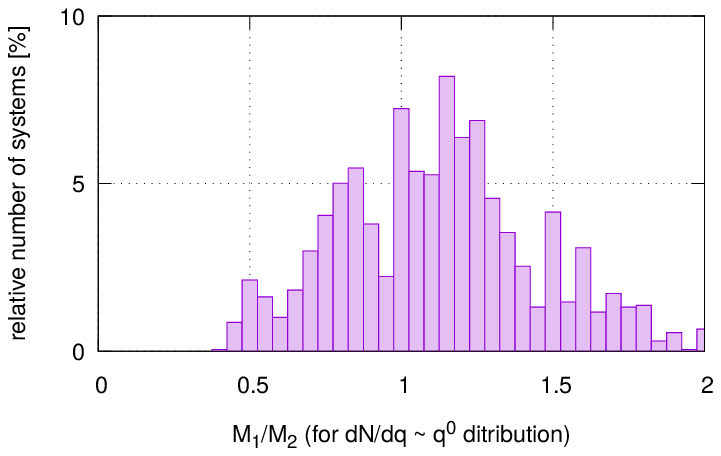} \\ a)
\end{minipage}
 \hfill
 \begin{minipage}[h]{1\linewidth}
\includegraphics[width=0.7\linewidth]{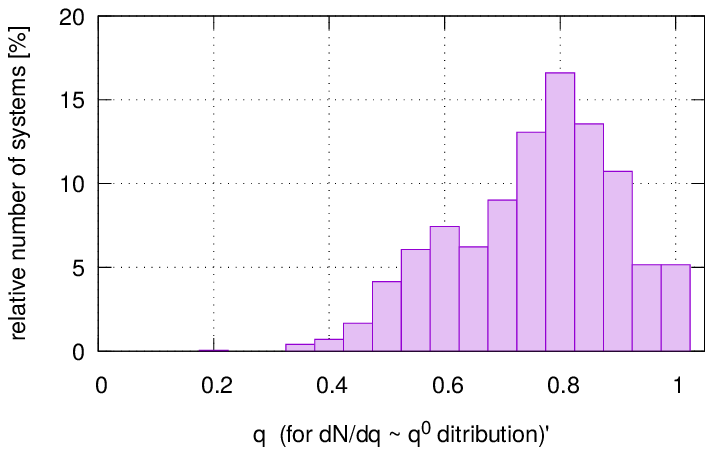} \\ b)
\end{minipage}
\end{figure}

\begin{figure}
\hfill
\begin{minipage}[h]{1\linewidth}
\includegraphics[width=0.7\linewidth]{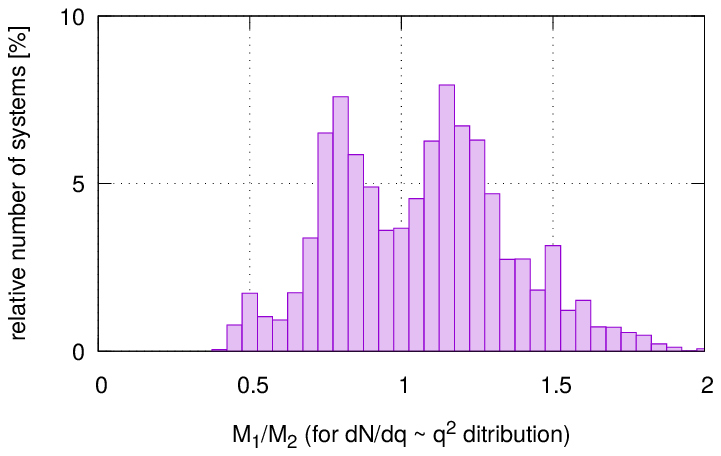} \\c)
\end{minipage}
\hfill
\begin{minipage}[h]{1\linewidth}
\includegraphics[width=0.7\linewidth]{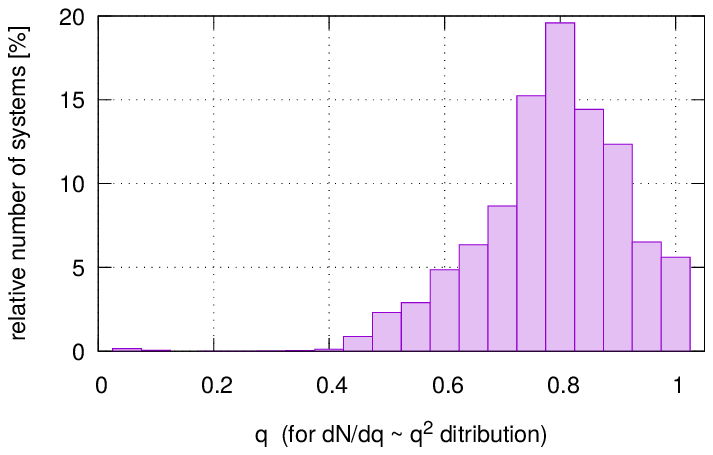} \\d)
\end{minipage}

\caption{ Distribution of the mass ratio  of merging black holes in the weak stellar wind scenario for initial mass of ratio distribution of  main sequence stars $\varphi(q_0) \sim q^{2}$  . Figure $a$ shows the distribution ratio $M_1 / M_2$ no matter which component is more massive at the time of the merger and where $M_1$ is the more massive star at  birth.  Figure $b$ shows $q$ distribution of merging black holes, where $q = M_{Light} / M_{Heavy}$. The $q$ - distribution has a single peak at $q \sim 0.8$, which is in good agreement with GW150914 ($q_{GW150914} = 29 M_{\odot} / 36 M_{\odot} = 0.8$).
This peak  splits into two for the   $M_1 / M_2$  distribution,  due to the fact that physically it is implemented in two different scenarios. The  two peaks are located at $M_1 / M_2  = 0.8 $ and  $M_1 / M_2  =  1.2 \sim 1/0.8 $. These scenarios differ in that in the latter case the secondary, having evolved to the helium-star stage, fills its Roche lobe and loses a substantial fraction of its mass before the supernova explosion. There is also  a third scenario, a scarcely populated domain which corresponds to large black-hole mass and high mass ratio. This is the case where the two components differed very much initially. There are the diagrams for $\varphi (q_0) \sim q^2 $  at $a)$ and $b)$ and for flat $\varphi 1 (q_0) \sim q^0 \sim const $ at $c)$ and $d)$ figures. These figures shows that the initial $q_0$ distribution hardly affects on the double black holes masses ratio distribution }
\label{figureDistrib2}
\end{figure}
\newpage

\begin{figure*}
\includegraphics[width=1\linewidth]
{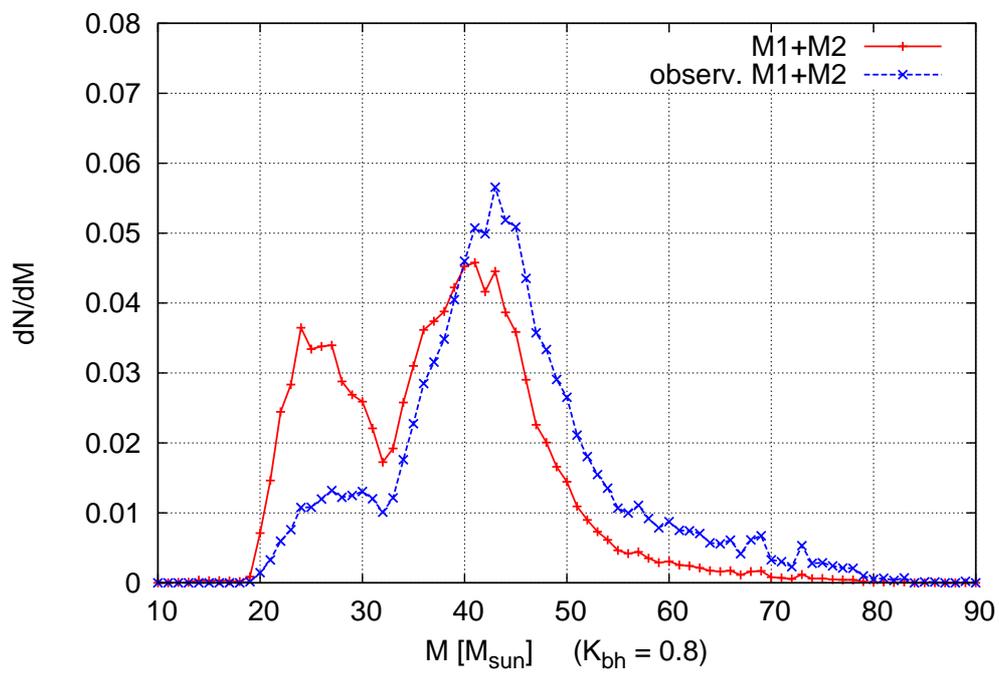}
\caption{ Distribution of the total mass of merging black holes in present Universe and the corresponding distribution for the events to be recorded by the detector.}
\label{figureDistrib}
\end{figure*}

\end{document}